\documentclass[]{aa}  

\usepackage{graphicx}
\usepackage{txfonts}
\usepackage{graphicx}
\usepackage{natbib}
\begin{document}
   \title{TW Hydrae: evidence of stellar spots instead of a Hot Jupiter \thanks{Based on observations taken at the VLT (Paranal), under programs 280.C-5064(A) and
     075.C-0202(A) , and with the CORALIE spectrograph
      and EulerCAM both at the Euler Swiss telescope (La Silla)}}
   \author{N. Hu\'elamo\inst{1}
         \and
         P. Figueira\inst{2}
         \and
         X. Bonfils\inst{3,4}
          \and
         N.C. Santos\inst{3}
         \and
          F. Pepe\inst{2}
        \and
         M. Gillon\inst{2}
         \and
         R. Azevedo\inst{3}
         \and
         T. Barman\inst{5}        
          \and
          M. Fern\'andez\inst{6}
          \and
          E. di Folco\inst{2}
          \and
         E.W. Guenther\inst{7}
         \and
         C. Lovis\inst{2} 
         \and
         C.H.F. Melo\inst{8}
         \and
         D. Queloz\inst{2}
          \and
         S. Udry\inst{2}
}

    \offprints{N. Hu\'elamo}
 \institute{Laboratorio de Astrof\'{\i}sica Espacial y F\'{\i}sica Fundamental 
 (LAEFF-INTA), Apdo. 78, E28691 Villanueva de la Ca\~nada, Spain \\
  \email{nhuelamo@laeff.inta.es}
   \and
   Observatoire de Gen\`eve, Universit\'e de Gen\`eve, 51, ch. des Maillettes, 
   1290 Sauverny, Switzerland
   \and
   Centro de Astrof\'{\i}sica, Universidade do Porto, Rua das Estrelas, 4150-762 Porto, Portugal
   \and
   Centro de Astronomia e Astrof\'{\i}sica da Universidade de Lisboa, 
   OAL, Tapada da Ajuda, 1349-018
   Lisboa, Portugal 
   \and
    Lowell Observatory, 1400 W. Mars Hill Rd., Flagstaff, AZ 86001
   \and
    Instituto de Astrof\'{\i}sica de Andaluc\'{\i}a, CSIC, Apdo. 3004,
    E-18080, Granada, Spain    
         \and
   Th\"uringer Landessternwarte Tautenburg, Karl-Schwarzschild-Observatorium, 
  Sternwarte 5, D-07778 Tautenburg, Germany
  \and
   European Southern Observatory, Karl-Schwarzschild-Strasse 2, D-85748 
   Garching bei M\"unchen, Germany   
   }
   \date{Received; accepted}

  \abstract
   {TW Hya is a classical T Tauri star that shows
   significant radial-velocity variations in the optical regime.
   These variations have been attributed to a 10 M$_\mathrm{Jup}$ 
   planet orbiting the star at 0.04\,AU. }
   {The aim of this letter is to confirm the presence of the giant planet
   around TW Hya by (i) testing whether the observed RV variations can be caused
   by stellar spots and (ii) analyzing new optical and infrared data to         
   detect the signal of the planet companion. }
   {We fitted the RV variations of TW Hya using a cool spot model.
   In addition, we obtained new high-resolution optical \& 
   infrared spectra, together with optical photometry of TW Hya
   and compared them with previous data.
   }
  {Our  model shows that a cold spot covering 7\% of the stellar surface
   and located at a latitude of 54$^\circ$ can reproduce the reported RV variations.
   The model also predicts a
   bisector semi-amplitude variation $<$ 10 m/s, which is
   less than the errors of the RV measurements discussed in Setiawan
   et al. (2008). 
   The analysis of our new optical RV data, with typical errors of 10 m/s,
   shows a larger RV amplitude that varies depending on the correlation mask used. 
   A slight correlation between the RV variation and the bisector is also observed
   although not at a very significant level.
   The infrared H-band RV curve is almost flat, showing a small variation ($<$35 m/s) that is
   not consistent with the published optical orbit. 
   All these results support the spot scenario rather than the presence of a hot Jupiter.
   Finally, the photometric data shows a 20\% (peak to peak)  variability, which is much larger
   than the 4\% variation expected for the modeled cool spot.
   The fact that the optical data are correlated with 
   the surface of the cross-correlation function points towards hot spots as being responsible for 
   the photometric variability.
   }
   {We conclude that the best explanation for the RV signal observed in TW\,Hya
   is the presence of a cool stellar spot and not an
   orbiting hot Jupiter.
   }

     
   \maketitle
%

\section{Introduction}

TW Hya (spectral type K7, M$_*$ =0.7$M_{\sun}$, age$\sim$10\,Myr) 
is an extensively studied classical T Tauri star (CTTS)
\citep[e.g.][]{1983A&A...121..217R}. 
With a distance of 56$\pm$7 pc
\citep{Esa1997}, it is probably one of the closest
laboratories for studying of our own solar system in its early stages
of formation. 

Recently, \citet[][SHL08, hereafter]{Seti2008} announced the discovery of a giant planet ($\sim$10\,M$_\mathrm{Jup}$) orbiting TW Hya at a separation of
0.04\,AU.  Their discovery is based on high-resolution optical
spectroscopic observations:  the analysis of two independent datasets
obtained with a difference of $\sim$ 3 months shows a 
significant radial velocity (RV, hereafter) variation with an
amplitude of 196$\pm$61 m/s and a periodicity of 3.56$\sim$0.02 days.  This,
together with the lack of correlation between the RV variation and the
cross correlation-function bisector (BIS), has been interpreted as proof
of the existence of a planetary-mass object orbiting the star.  

In this letter, we study whether the RV variations in TW Hya can be
caused by stellar spots and not by a planetary-mass companion.  With this
in mind, we have tried to reproduce SHL08 data with a cool spot model. 
In addition, we obtained almost simultaneous optical \& near-infrared high resolution spectroscopy and optical photometry of the source in order to detect the planet signal with a different dataset.

\section{A spot model for TW Hya}
\label{sec:spot}


To test whether a spotted surface is a viable model for explaining the SHL08 dataset,
we simulated the changes of TW Hya's line profile due to such a
spot. For that purpose, we used \textsc{Soap} (Bonfils \& Santos 2008,
in prep.), a program that calculates the photometric,
radial-velocity, and line-shape modulations induced by one (or
more) cool stellar spot.

\begin{figure}\label{modelSpot}
\centering
\resizebox{\hsize}{!}{
\includegraphics[width=0.15\textwidth]{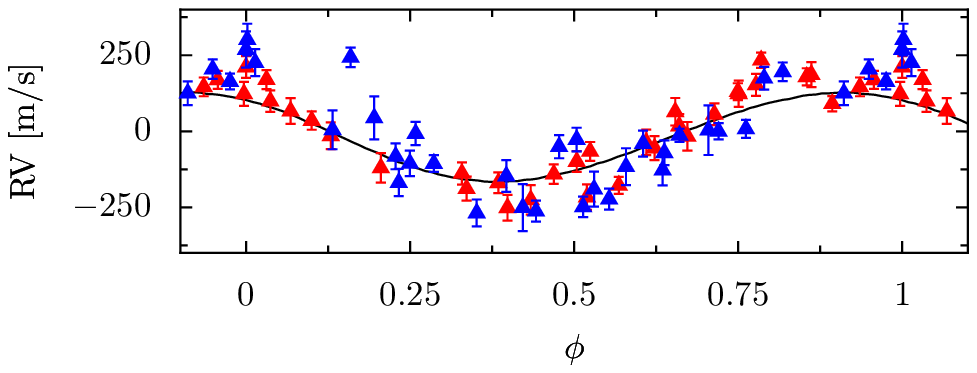}}
\resizebox{\hsize}{!}{
\includegraphics[width=0.15\textwidth]{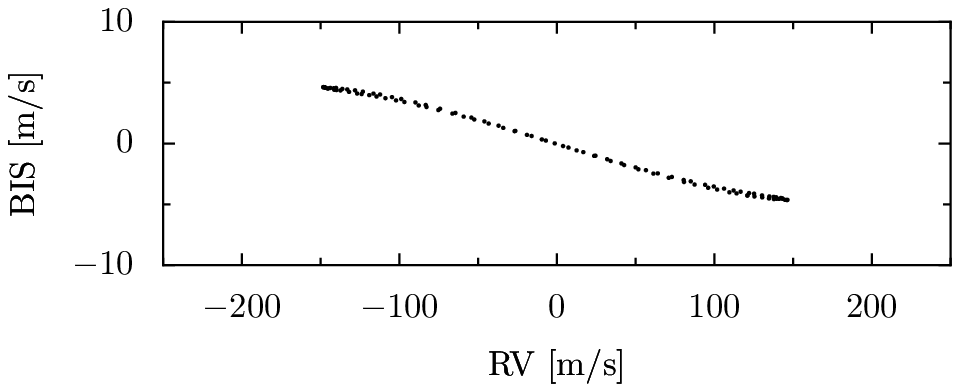}}
\caption{Radial velocity and bisector variations as calculated by our spot model. In the top panel, the
  two epoch (clear and dark symbols) RV measurements from Setiawan et al. (2008) are  superimposed.}
  \vspace{-0.2cm}
\end{figure}

\textsc{Soap} computes the rotational broadening
of a spectral line by sampling the stellar disk on a grid. For each
grid cell, a Gaussian function represents the typical line of the
emergent spectra.  It has the same width, $\sigma_0$, as the typical
spectral line of a star with no rotational velocity (or one too low to be
resolved) and of similar spectral type than the star of interest. The
Gaussian is Doppler-shifted according to the projected rotational
velocity ($v \sin i$) and weighted by a linear limb-darkening law
($\alpha=0.6$). For a given spot (defined by its latitude, longitude, 
and size), \textsc{Soap} computes which of the grid cells are obscured
and removes their contribution to the (integrated) stellar spectrum.
Finally, the stellar spectrum output by \textsc{Soap} is the sum of
all contributions from all grid cells. In our case, the stellar
spectrum is an averaged spectral line. It behaves like the
cross-correlation profile of stellar spectra, and \textsc{Soap} uses it
to compute the RV and the bisector, in addition to the
photometric variation.

In the case of TW Hya, we chose
$R_\star=0.9~\mathrm{R_\odot}$ (following SHL08)
and an inclination of  $i=7^\circ$, which is the likely inclination of the TW Hya circumstellar disk
\citep{qi2004}.
The strong veiling in the atmosphere of TW Hya prevents us from using its B$-$V color
index as a temperature proxy. Both $\sigma_0$ and $v \sin i$ 
will probably yield wrong estimates if derived from a calibration based on
this color index. Therefore, we only used the B-V color 
to derive a first estimation of $\sigma_0$$= 4.45~\mathrm{km\,s^{-1}}$ \citep{melo2001}.
We then iterated on different values to show
that our conclusions are unchanged. The rotational period of the star, the
spot size and latitude, and the star velocity 
($\gamma$) were left as
free parameters. Adjusting these parameters, we were able to reproduce
SHL08's RVs with a dark spot covering about 7\% of the stellar
surface, located at a latitude of 54$^\circ$ and with
a rotation period of 3.56 days. With the assumed stellar radius and inclination,
and with the fitted rotational period, we derived a $v.sini$ of 1.56  $\mathrm{km\,s^{-1}}$.

Remarkably, our solution has
$\chi^2=3.27$, very similar to the Keplerian solution
derived by  SHL08 ($\chi^2$=3.32). 
Our model also indicates that such a spot would
produce a 4\%  photometric variability in the V-band, and a peak-to-peak amplitude
change $<$10 $\mathrm{m\,s^{-1}}$ in the BIS measurements (see Fig. 1). An extreme
choice of $\sigma_0$ =1  $\mathrm{km\,s^{-1}}$ increases the
amplitude of the BIS variation up to $<$30 $\mathrm{m\,s^{-1}}$.


SHL08 measured the BIS and conclude that the observed RV change is not due
to a spot. However, their BIS measurements have individual errors of 
$\sim$50-100 m/s and their overall dispersion is almost as large as the RV
variation, therefore much higher than the variation expected from our
model. Varying the constraint on the inclination, we find solutions with
$\chi^2 $$<$3.6 from $i=2^\circ$ to $i\sim80^\circ$. 
With $\sigma_0$=4.55 $\mathrm{km\,s^{-1}}$, the peak-to-peak variations of 
BIS were found to be lower
than $100 \,\mathrm{m\,s^{-1}}$ for all inclinations below $\sim$ 50$^\circ$.%

According to our model, the low inclination of TW Hya's 
system weakens significantly the
power of a bisector analysis and other diagnostics should be used to
validate or disprove the planet hypothesis.

\section{New observations of TW Hya}

\subsection{High-resolution optical spectroscopy}

With the goal of confirming the period, phase, and amplitude of the
RV signal observed by SHL08, we used the CORALIE
spectrograph (1.2-m Euler Swiss telescope, La Silla, Chile) to obtain
a series of 12 consecutive radial-velocity measurements of TW\,Hya. The data
were gathered between February and March 2008. Each measurement has an
accuracy of about 10\,m\,s$^{-1}$. CORALIE has a spectral
resolution R=$\lambda/\Delta\lambda\sim50\,000$, and is known to have
a long term precision in radial-velocity of a few
$\sim$3\,m\,s$^{-1}$ \citep[e.g.][]{Mayor-2004,Wilson2008}.

The CORALIE RVs are obtained by cross-correlating the
stellar spectrum with an optimized weighted template 
\citep[for details see ][]{Pepe2002}. For TW\,Hya, we performed 
correlations using 
cross-correlation masks that are
appropriate for stars with different spectral types: G2, K0, K5 (the one
closer to the spectral type of TW\,Hya), and M4. This allows to verify
that the amplitude of any putative RV signal shows the same
value, independently of the mask used. This is expected if
the radial-velocity variations observed stem from the presence of a
planet. On the other hand, an amplitude dependence on the mask used
could be expected if the RV signal is produced by the
presence of a stellar blend \citep[][]{Santos-2002a} or a cool stellar
spot. In the latter case, the correlation with a mask built for
a later spectral type (e.g. M4) should provide a signal with a smaller
amplitude than the one derived using a mask for an earlier
spectral type (e.g. G2). This is due to the fact that later type masks
are more sensitive to the lines in the spot than are the earlier type masks.
In the former case, the spectroscopic contrast between the spot and
the rest of the photosphere is thus weaker.


  \begin{figure}
   \centering
      \resizebox{\hsize}{!}{\includegraphics{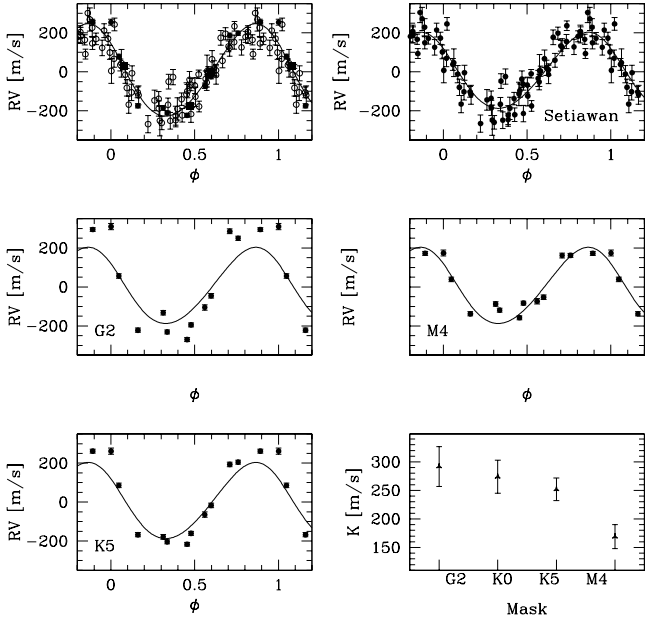}}
      \caption{{\it Top left}: Phase-folded radial-velocity
        measurements of TW\,Hya with the Keplerian fit obtained
        with the HARPS (squares), CORALIE (filled circles), and
        FEROS (open circles) data.  The best Keplerian
        fit to the three groups of points is shown.  The
        semi-amplitude of this fit is K=238\,m\,s$^{-1}$.  {\it Top
          right}: Original data from SHL08. The best
        Keplerian fit is shown (K=196\,m\,s$^{-1}$).  {\it Mid and
          lower left panels}: CORALIE phase-folded radial-velocities
        derived using three different CCF masks (see text for  details). 
        The same Keplerian function as found using SHL08 
        is shown for comparison.  {\it Lower right}:
        RV amplitude of the best Keplerian fit to the
        CORALIE data derived using 4 different masks.}
   \label{fig:vr}
   \vspace{-0.05cm}
   \end{figure}

The CORALIE RVs have been complemented by older
RVs measurements of TW\,Hya (three datapoints) obtained with the HARPS 
spectrograph \citep[data described in][]{Guenther2007}, 
and the FEROS RVs listed by SHL08.
Using the HARPS, FEROS, and CORALIE (K5-mask)
radial-velocities together, we find a clear signal with a
period of $\sim$3.56\,days, and a semiamplitude of
K=238$\pm$9\,m\,s$^{-1}$ (Fig.\,\ref{fig:vr}, upper-right panel). The
parameters of a Keplerian fit are presented in
Table\,\ref{tab:elements}. 
Using only the points listed by SHL08 and our own fitting routines, we do confirm their solution, with a semi-amplitude of K=196$\pm$13\,m\,s$^{-1}$. If 
caused by the presence of a
planet, our data could thus imply a planetary mass that is 20\% higher
than the one announced by SHL08.

Interestingly, however, a look at the CORALIE measurements alone shows that by using different cross-correlation masks we can derive significantly different amplitudes when fitting a Keplerian function to the data (Fig.\,\ref{fig:vr} ; bottom-right panel).  For instance, using a mask built for a G2-type
dwarf we derive a value of K=292$\pm$35\,m\,s$^{-1}$.  This value
decreases to K=274$\pm$29\,m\,s$^{-1}$, K=252$\pm$20\,m\,s$^{-1}$, and
K=169$\pm$21\,m\,s$^{-1}$, using the K0, K5, and M4 masks,
respectively\footnote{To derive these values we fixed the eccentricity
to zero. No major differences were found if the eccentricity is set
as a free parameter.}. This is in line with the presence of a spot instead of
a planet to explain the observed RV variations.
  
If the signal is caused by a stellar spot, a color effect can also
be expected. If we correlate the spectra with a given mask using only the blue or the red sides,  we can expect to obtain slightly different values for the RV amplitude. This difference comes from the different contrasts (flux ratio) coming from the spot and the surrounding photosphere as a function
of wavelength. In a simple blackbody approximation, the contrast between a
spot and the surrounding disk is stronger at bluer wavelengths, thus implying a stronger RV signal. 

To test this, we derived CORALIE RVs using the K5 mask but correlating 
only a red portion of the spectrum (4150--4850\AA)
and a blue portion of the spectrum (6000--6700\AA). In the former case, 
we obtained a semi-amplitude of K=228$\pm$25\,m\,s$^{-1}$, while in the latter  we derived K=210$\pm$20\,m\,s$^{-1}$. These values are marginally compatible
with a small variation. Indeed, a simple comparison of the effect due to the different flux ratios in the different bands used (considering a spot with a temperature around 3000\,K and the TW\,Hya temperature of 4000\,K -- SHL08) shows that a variation of about 10\% can be expected. 
Following this rationale, the RV amplitude in the infrared (IR) regime is expected to be much smaller ($\sim$5 times smaller) if the RV signal is indeed caused by a spot.  We test this hypothesis in Sect. 3.2.

\begin{figure}[t]
\centering
\includegraphics[width=0.35\textwidth]{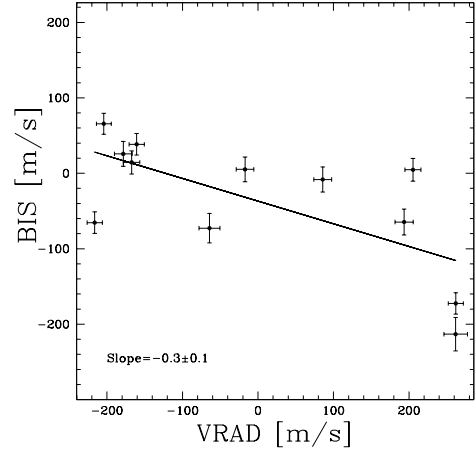}
\caption{BIS vs. radial-velocity (derived with the K5 mask) for the
  CORALIE data of TW\,Hya.  For clarity, the vertical
  and horizontal scales were set to be the same.  The best fit to the
  data is also shown.}
\label{fig:bis}
\end{figure}
\begin{table}[h!]
\caption[]{Elements of the fitted Keplerian function.}
\begin{tabular}{lll}
\hline
\hline
\noalign{\smallskip}
$P$             & 3.5683$\pm$0.0002			& [d]\\
$T$             & 2454530.1$\pm$0.3		& [d]\\
$e$             & 0.07$\pm$0.04			&  \\
$V_r$$\dagger$ (HARPS) & 12.498$\pm$0.016	& [km\,s$^{-1}$]\\
$V_r$$\dagger$ (FEROS) & 12.4204$\pm$0.011	& [km\,s$^{-1}$]\\
$V_r$$\dagger$ (CORALIE) & 12.518$\pm$0.008	& [km\,s$^{-1}$]\\
$\omega$        & 55$\pm$29			& [degr] \\ 
$K_1$           & 238$\pm$9			& [m\,s$^{-1}$] \\
$\sigma(O-C)$   & 49				& [m\,s$^{-1}$]  \\    
\noalign{\smallskip}
\hline
\end{tabular}
\label{tab:elements}
\end{table}

As explained in Sect.~2, the 
absence of any clear correlation between the RV variation and the 
bisector of the cross-correlation function was considered by the SHL08 as 
evidence of a planet orbiting TW\,Hya. However,
the errors in their data (around 100\,m\,s$^{-1}$) may have
hidden any low amplitude signal. To test this, we computed the BIS values for
our CORALIE measurements \citep[for the definition see][]{Queloz-2001}. 
As can be seen in Fig.\,\ref{fig:bis},  the BIS and RV are slightly correlated  (slope=$-$0.3$\pm$0.1), 
though not at a very significant level (correlation coefficient, $r$=0.82). This result is compatible with the output from the simple model described in Sect.\,\ref{sec:spot}.

Finally, we searched for phase variations in the RV signal between the FEROS (SHL08) and CORALIE data, obtained with a time span of about one year. The results show that no measurable phase shift was found over more than 300 days to a precision of~1\%.

\subsection{High resolution near-IR spectroscopy}

A planet orbiting TW Hya should be detectable
with a similar RV amplitude and orbital period at all wavelength ranges.
As mentioned above, if the RV signal reported by SHL08 
is caused by a cool spot, the RV amplitude is expected to be 
smaller in the IR regime.
To test these two scenarios, we observed TW Hya with 
CRIRES, the CRyogenic high-resolution InfraRed Echelle Spectrograph mounted on the VLT \citep{Kauf2006}.  To obtain the most accurate RV measurements, we selected a specific setting in the H band defined  as 36/1/n   ($\lambda_c$= 1.5884$\mu$m; $\Delta$$\lambda$$\sim$26$\mu$m)  in the CRIRES manual. 
This setting is populated by
deep and sharp CO$_{2}$ lines that can be used as a simultaneous
wavelength reference. 
The observations were done with a slit width of
0.2", using a typical ABBA nodding sequence, and a seeing $>$ 0.8" in the optical -- the objective was to reduce photocenter-induced RV errors while observing
at the highest resolution available, estimated as 100\,000.

To derive the RV from the spectra we used a
cross-correlation method \citep{baranne1996}. The spectra were
correlated simultaneously with a telluric mask, built with the HITRAN
database \citep{Rothman1998}, and a stellar mask, built from PHOENIX
models \citep{Barman2005}.  The atmospheric lines allow us to
establish the zeropoint of the wavelength calibration. By subtracting
the RV of the atmospheric lines from the target RV, we can correct
from instrumental changes following the same principle of the
simultaneous ThAr, \citep[e.g.][]{Pepe2007}.  The whole procedure will
be described in detail in a forthcoming paper (Figueira et al., {\it in prep.}).

\begin{figure}[t]
\begin{center}
\parbox{9.5cm}{
\vspace{-0.3cm}
\centering
\resizebox{9cm}{!}{\includegraphics{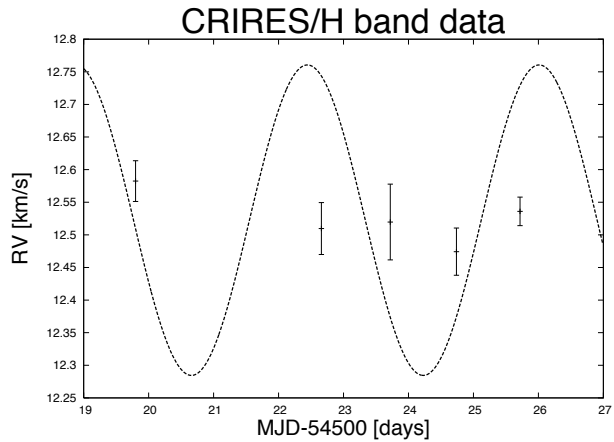}}
}
\caption{CRIRES/H-band Radial-velocity points overplotted on the 
fitted Keplerian function shown in Table~1. The error bars correspond to the quadratic addition of the errors bars on the RV standard and on TW  Hya measurements.
}
\label{fig:CRIRES}
\end{center}
\end{figure}

TW Hya was observed five times between Feb 22 and 28 2008.  
An RV standard (HD108309) was observed immediately after the target each night.
We calculated the mean RV per night and assigned
a 1$\sigma$ error bar corresponding to the standard deviation of the
RV in the nodding cycle. We verified that the measured scatter per night, between 20 and 50 m/s, matched the expected uncertainties as estimated by \citet{Bouchy2001}. However, the scatter
of the 5 datapoints obtained on different nights is higher ($\sim$\,90\,m/s), testifying the presence of systematic errors. Among others, all the observations in the first night were done with
the guiding camera saturated. To remove these effects, and assuming
that they affect both TW Hya and the RV standard in the same way, we
calculated the relative velocity between TW Hya and the standard for
each night. The rms of the five relative RV is 35 m/s, and 22 m/s if
we only consider the 4 last points. The results are plotted in
Fig.~\ref{fig:CRIRES} and compared with the fitted orbit (Table~\ref{tab:elements}).

To estimate the match between the IR observations
and the fitted orbit (Table~\ref{tab:elements}), we performed a Monte Carlo simulation. For each JD, 
we built an RV point by adding to the corresponding orbit value an
observational error represented by a zero-centered Gaussian
distribution with a fixed dispersion. We calculated the probability of
obtaining the observed RV dispersion, hence the probability that our five
points indeed belong to the fitted orbit. 
For the observed scatter of 35 m/s (the measured one) and 70 m/s (twice the observed one), the probability that our RV dispersion is drawn from the fitted orbit is always below 10$^{-6}$.

Finally, we note that, as an important by-product of this work,  we have obtained the most accurate RV 
measurements in the near-IR to date,  showing the capabilities of CRIRES in this domain.

\subsection{Photometric observations}

\begin{figure}
   \centering
    \vspace{-0.5cm}
     \includegraphics[width=0.5\textwidth]{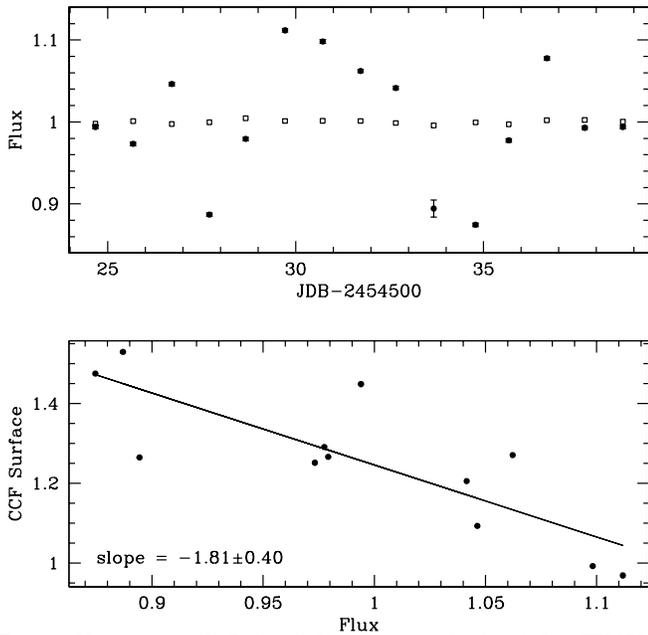}
    \vspace{-1.0cm}
     \caption{{\it Upper panel}: R-band lightcurve obtained for TW\,Hya
        (filled symbols) and for the main reference
        star (open symbols).  The fluxes are normalized. 
        {\it Lower panel}: TW\,Hya
        photometric flux as a function of the surface of the CORALIE
        CCF. The best linear fit to the data is shown.  }\label{twhphot}
   \end{figure}

TW Hya was monitored photometrically with the Euler telescope (La
Silla, Chile) during 15 nights from 2008 February 2 to March
13 under photometric conditions (except for March 8). Each night, a sequence of six exposures was observed at approximatively the same airmass to minimize any second-order
extinction effect. Observations were performed in the R-band and were
defocused to $\sim$ 5\arcsec to (1) minimize any systematics due to
the intra/inter-pixels variations and (2) to allow exposure times of
$\sim$ 50s in order to minimize the scintillation noise. The pointing
was chosen to include TYC\,7208-1066-1 and 3 other fainter
isolated comparison stars in the EulerCAM field of view. 
Aperture photometry was used to extract 
the fluxes. Figure~\ref{twhphot} shows the resulting lightcurve. 

TW Hya exhibited $\sim$20\% (peak to peak) flux variations during the
run.  The analysis of the lightcurve using both the Lomb-Scargle
periodogram \citep{scarg1982} and the string-length method
\citep{do1983} provides tentative periods of 6.1 and 6.5 days,
respectively.  The origin of this period is not clear and the small
time coverage does not allow us to draw further conclusions about this
result.  The comparison of this value with previous estimates does not
help in interpreting our data, since the rotation period of TW Hya is
not well-established and very different values (from 1.6 to 4.4 days)
have been reported in the literature
\citep[e.g. ][]{Mekka1998,Batalha2002,lawson2005,Seti2008}.

Interestingly, the photometric variations seem to be correlated with
the surface of the CORALIE CCF (the CORALIE data and the photometric
measurements were obtained on the same nights) --
Fig\,\ref{twhphot}, lower panel.  Since the CORALIE CCF is a
measurement of an average spectral line in the spectral region used,
the measurement of its surface is related to the veiling due to
the hot spots on the stellar surface \citep[e.g.][]{Andersen1989}. 
The correlation displayed  in Fig. \,\ref{twhphot} would imply that the photometric variations of TW Hya are 
dominated  by hot spots, something that has already been reported
in several studies \citep[e.g.][]{Mekka1998, Batalha2002}.
The V-band amplitudes  measured in these works range between  0.2-0.5\,mag, that is, they are much larger
than the 4\% V-band photometric variations predicted by our cold-spot model.
The fact that the photometric lightcurve cannot be fitted 
with the 3.56 days period derived from the spectroscopic data
(and assuming this as the true rotational period) can be related to
the presence of short-lived hot spots (with typical lifetimes of hours or days) 
that appear and disappear hiding the true period of the source \citep[see][]{Mekka1998}.

We note that, although the effects of a hot spot on the RV
variations are not tested here, we expect them to be significantly
weaker than the ones induced by a cold spot. Note that their
filling factors are usually between 0.1\% and 5\%
\citep[e.g.][]{Mati1996}, that is, relatively small compared to the
one of the dark spot from the model described in Sect.~\ref{sec:spot}. 

\section{Conclusions}

The analysis of new optical and infrared spectra of TW Hya  
allows us to conclude that the best explanation
for the observed RV signal is
the presence of a cold spot and not a  hot Jupiter orbiting the star. 
First, we find a clear dependence of the 
optical RV amplitude on the CCF mask used 
and, second, the IR RV curve is almost
flat, with a small RV scatter ($\sim$35 m/s) that is inconsistent with
the optical, orbital solution. In fact, we show that a simple cool
spot model can reproduce the RV observations presented by
SHL08. On the other hand, the photometric data are
correlated with the surface of the CCF, which points towards a hot spot as
the cause of the photometric variability, as reported in previous studies.


Our result shows that searching for planets around young stars is highly 
demanding and cumbersome. Only the concerted use of the different techniques and instruments presented here were able to clarify the nature of the signal observed in TW\,Hya.
It is, for instance, remarkable that no phase shifts were found in the RV signal between  different epochs (Sect. 3.1), which shows that very stable cool spots exist around CTTSs. This must be taken with care when searching for planets around this kind of objects.

\begin{acknowledgements}
 We are grateful to the  Paranal Science Operation team, in particular
 to A. Smette. We thank the ESO Director Office for granting DDT
 observations.  NH is funded by Spanish grants
 MEC/ESP 2007-65475-C02-02, and
 CAM/PRICIT-S-0505/ESP/0361. NCS, XB, and PF acknowledge the support from Funda\c{c}\~ao para a Ci\^encia e a Tecnologia (FCT), Portugal, in
 the form of grant PPCDT/CTE-AST/56453/2004 and  scholarship SFRH/BD/21502/2005. XB acknowledges the Gulbenkian  Funda\c{c}\~ao for support through its {\em Est\'{\i}mulo da Investiga\c{c}\~ao} program. 
 M.F. was supported by the Spanish grants AYA2006-27002-E, and AYA2007-64052.
  \end{acknowledgements}

\bibliographystyle{aa}
\bibliography{0596}

\end{document}